\documentclass[9pt,twocolumn,twoside]{opticajnl}
\journal{opticajournal} 

\setboolean{shortarticle}{true}

\usepackage{lineno}

\usepackage[algo2e]{algorithm2e}
\usepackage{subfigure}

\title{Generation of electro-optic frequency combs with optimized flatness in a silicon ring resonator modulator}

\author[1,5,*]{Erwan Weckenmann}
\author[2]{Thyago Monteiro}
\author[2]{Uiara Celine de Moura}
\author[2]{Francesco Da Ros}
\author[1]{Laurent Bramerie}
\author[1]{Mathilde Gay}
\author[3,6]{Diego Pérez-Galacho}
\author[1,3]{Lucas Deniel}
\author[4]{Frédéric Boeuf}
\author[3]{Delphine Marris-Morini}
\author[2]{Darko Zibar}
\author[1]{Christophe Peucheret}

\affil[1]{Univ Rennes, CNRS, FOTON -- UMR 6082, 22305 Lannion, France.}
\affil[2]{Department of Electrical and Photonics Engineering, Technical University of Denmark, 2800 Kgs. Lyngby, Denmark.}
\affil[3]{Université Paris-Saclay, CNRS, Centre de Nanosciences et de Nanotechnologies, 91120 Palaiseau, France}
\affil[4]{ST Microelectronics, 850 rue Jean Monnet, 38920 Crolles, France}
\affil[5]{Department of Electrical and Computer Engineering, Centre for Optics, Photonics and Lasers, Université Laval, Quebec, QC G1V 0A6, Canada}
\affil[6]{Photonics and RF Research Lab, Universidad de Málaga, 29071 Málaga, Spain}
\affil[*]{Corresponding author: erwan.weckenmann.1@ulaval.ca}

\begin{abstract}
The flatness of electro-optic frequency combs (EOFCs) generated in a single silicon ring resonator modulator (RRM) is optimized by employing harmonic superposition of the radio-frequency driving signal. A differential evolution algorithm is employed in conjunction with a simplified model of the RRM for offline optimization of the amplitudes and phases of harmonic driving signals and the operating point of the RRM. The optimized driving signals are then applied to a silicon RRM. EOFCs containing 7 and 9 lines are synthesized with a power imbalance between the lines of 2.9~dB and 5.4~dB, respectively, compared to 9.4~dB for an optimized 5-line comb generated from a single sinusoidal driving signal.
\end{abstract}

\setboolean{displaycopyright}{false} 

\begin{document}

\maketitle

Optical frequency combs (OFCs) are being considered in many application fields such as spectroscopy, metrology or as multi-wavelength sources for optical communications or microwave photonics \cite{Diddams_science_20_v369_peaay3676}. Depending on the targeted number of lines, line spacing, center wavelength and power levels, different comb generation mechanisms have been proposed and demonstrated \cite{Diddams_science_20_v369_peaay3676}. Among those, electro-optic frequency combs (EOFCs), where the comb is generated by externally modulating a continuous wave (CW) laser, present advantages in terms of centre wavelength and line spacing tunability \cite{Torres-Company_lpr_14_v8_p368}. Such combs have been studied or demonstrated using a variety of electro-optic modulators based on bulk LiNbO$_3$ or InP. Those include phase modulators (PMs), Mach-Zehnder modulators (MZMs) \cite{Sakamoto_ol_08_v33_p890,Yokota_ol_16_v41_p1026,Yokota_jqe_16_v52_p5200207,Ellis_ptl_05_v17_p504}, in-phase and quadrature (IQ) modulators \cite{Wang_ol_14_v39_p3050} as well as cascades of those structures.

Electro-optic modulators that can be integrated on the silicon photonic platform are being intensively investigated for applications where reduced footprint, cost and energy consumption are critical, for instance in short-reach data links, microwave photonic interfaces, on-chip optical networks and photonic computing applications. In these applications, minimizing the number of laser sources could present an advantage, making the synthesis of EOFCs with silicon modulators particularly relevant. Silicon-modulator combs also have potential for other applications than those related to data or telecommunications, for instance dual-comb spectroscopy~\cite{Deniel_ox_20_v28_p10888}.

Among silicon photonic compatible modulators, those directly relying on carrier-density modulation in silicon benefit from technical maturity with continuous progress in terms of bandwidth and modulation efficiency. EOFC generation has been demonstrated in carrier depletion or injection-based silicon modulators using different structures such as a single phase modulator~\cite{Nagarjun_ox_18_v26_p10744}, a single MZM~\cite{Lin_ptl_18_v30_p1495,Deniel_pres_21_v9_p2068}, cascaded MZMs~\cite{Wang_jqe_19_v55_p8400206,Liu_jstqe_20_v26_p8300208}, parallel MZMs~\cite{Liu_jlt_20_v38_p2134} or cascaded PM-MZM structures \cite{Deniel_pres_21_v9_p2068}. The use of loop-synchronous phase modulation has also been reported in a lithium-niobate-on-insulator resonant structure \cite{Zhang_nature_19_v568_p373}.

One widely-investigated silicon modulator structure is the ring-resonator modulator (RRM), which is attractive from a compactness and power efficiency point of view. Some combs have been demonstrated using single RRMs \cite{Wu_cleo_17_SM4O.6,Demirtzioglou_ox_18_v26_p790}, however, with a modest number of lines. Cascaded RRMs can then be used to further broaden the comb \cite{Xu_ol_18_v43_p1554,Nagarjun_ox_20_v28_p13032,Khalil_iprsn_20_IM3A.6}. One open question is therefore to assess the potential of these modulators for EOFC generation and, in particular, optimize the number of lines that can be obtained with acceptable power imbalance from a single RRM.

In this Letter, we report a flatness optimization for EOFCs synthesized using a single carrier-depletion RRM. The offline optimization procedure relies on a differential evolution algorithm employed in conjunction with a numerical model of a silicon RRM. Such an algorithm has already been employed for the numerical optimization of EOFCs generated from MZMs with linear phase-shifters~\cite{Pendiuk_osac_20_v3_p2232}. Here, a joint offline optimization of the amplitudes and phases of a sum of three harmonics used to drive the modulator, as well as of the detuning between the resonance and the input CW laser, is performed in order to minimize the power imbalance between the lines of 7- and 9-line combs. The numerical optimization results are then validated experimentally by applying the numerically-optimised radio-frequency (RF) signal to an RRM. Using this method, a 7-line EOFC with a flatness (defined as the power imbalance among the specified number of lines) of 2.9~dB and a 9-line EOFC with a flatness of 5.4~dB are generated, each with a frequency spacing of 500~MHz. As detailed below, this spacing has been chosen to ease the comparison between numerical optimization and experimental demonstration, but is scalable when taking the electro-optic response of the modulator into account. Our results significantly improve the performance of single-RRM combs reported earlier in the literature in terms of number of lines and flatness.

The considered RRM is based on a standard all-pass ring configuration with a carrier-depletion based phase-shifter implemented as a lateral PN junction in the ring waveguide~\cite{Morini_ieeeproc_09_v97_p1199}. It has a ring radius of 20~$\mu$m and a loaded \textit{Q}-factor in the absence of applied voltage of $3~\times~10^4$, corresponding to an optical bandwidth of 6.5~GHz. In order to calculate the refractive index change as a function of applied voltage, a rib waveguides of 400-nm width, 150-nm etching depth for a 310-nm total silicon height was considered. The P-doped and N-doped regions have carrier concentrations of $5\times 10^{17}$~cm$^{-3}$ and $10^{18}$~cm$^{-3}$, respectively.

\begin{figure}[t]
    \centering
    \includegraphics[width = 0.85\columnwidth]{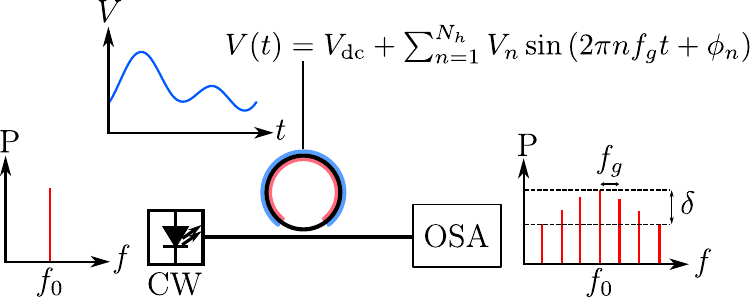}
    \caption{Principle of EOFC comb generation in an RRM. A sum of $N_h$ harmonics with fundamental modulation frequency $f_g$, amplitudes $V_n$ and phases $\phi_n$ is applied as a driving voltage to the phase-shifter of the RRM, resulting in a comb with line spacing $f_g$ centered on the optical carrier frequency $f_0$ of the input continuous-wave (CW) laser. The comb power imbalance $\delta$ is measured using an optical spectrum analyzer (OSA).}
    \label{fig:setup}
\end{figure}

Figure~\ref{fig:setup} represents the experimental setup, which consists in inputting CW light from a tunable laser source to the RRM and modulating it by applying an RF voltage to the phase-shifter. The source wavelength ($\lambda_0$) is tuned with respect to the resonance wavelength ($\lambda_r$) at a given direct-current (DC) bias voltage $V_{\mathrm{dc}}$
 according to a wavelength-detuning value defined as $\Delta\lambda = \lambda_0 - \lambda_r$. The spectrum of the output modulated electric field is then measured using a high-resolution (5-MHz) optical spectrum analyzer (OSA). The comb flatness is quantified by a power imbalance parameter ($\delta$, expressed in decibel) defined as the ratio between the maximum and minimum line power among a specified number of comb lines centered around the CW laser frequency $f_0$.

\begin{figure}[t]
    \centering
    \includegraphics[scale=0.3]{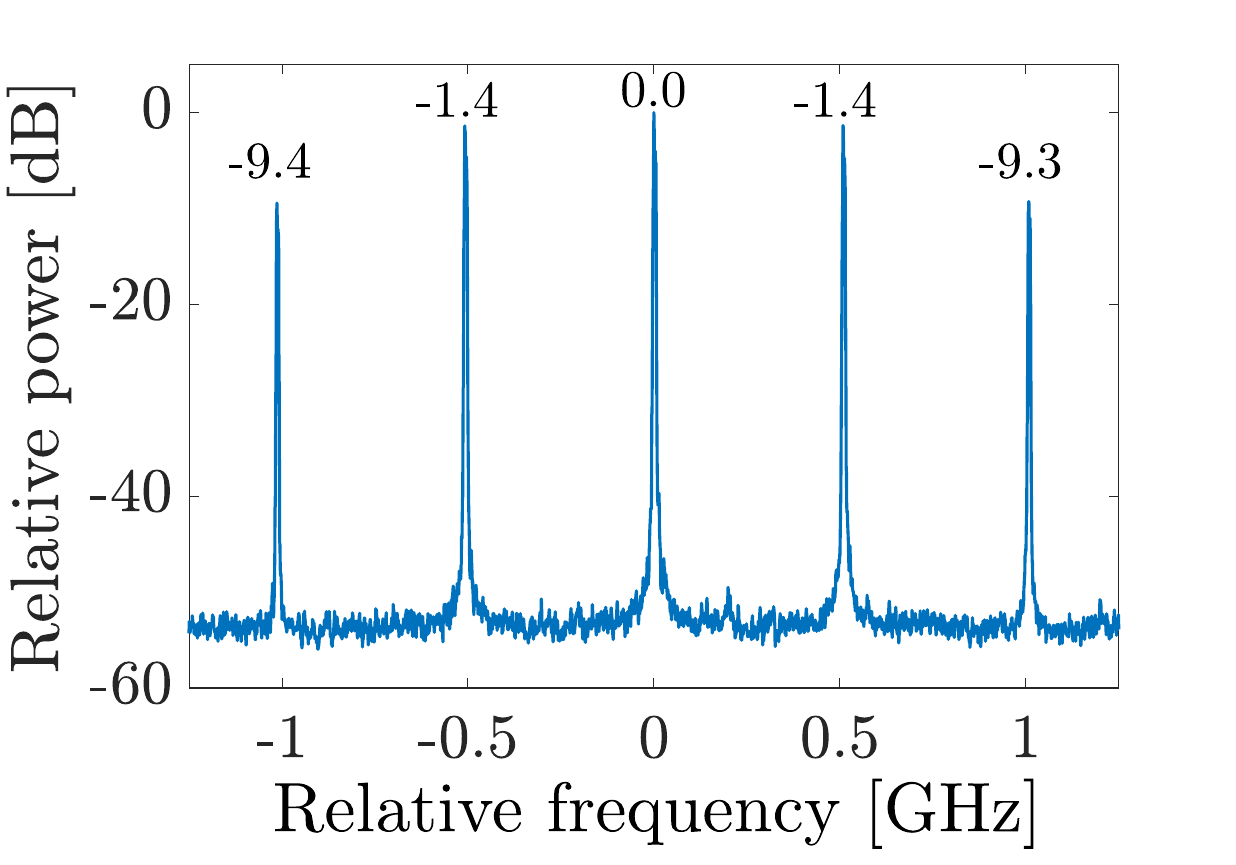}
    \caption{Measured comb spectrum for single-tone modulation at 500~MHz with a peak-to-peak amplitude of 6.4~V.}
    \label{fig:mes_sinusoidal_5lines}
\end{figure}

Figure~\ref{fig:mes_sinusoidal_5lines} shows a typical comb spectrum measured at the output of the silicon RRM when a single sinusoidal tone (frequency of 500~MHz and peak-to-peak voltage of 6.4~V) is applied to the phase-shifter. The wavelength detuning between the source and the resonance is set in this case to $\Delta\lambda = +10$~pm to minimize $\delta$ for 5 lines. The spectrum consists in several pairs of sidebands  spaced by the modulation frequency and symmetrically located around the optical carrier. A minimum power imbalance among 5~lines of $\delta = 9.4$~dB is obtained in this configuration.

The comb flatness $\delta$ can be improved by tailoring the modulator driving signal \cite{Yokota_ol_16_v41_p1026}. It was already shown in \cite{Demirtzioglou_ox_18_v26_p790} that the superposition of two harmonics in the driving signal of an RRM could lead to improvement in the power distribution of the comb lines. However, no systematic optimization was reported. In this work, we conduct such an optimization by considering an increased number of parameters, which is made tractable by the use of a meta-heuristic algorithm, more specifically a differential evolution (DE) algorithm  \cite{Storn_jgo_97_v11_p341}.

An RF signal composed of $N_h = 3$ superposed harmonics of a fundamental frequency $f_g$ is considered, as illustrated in Fig.~\ref{fig:setup}. Each harmonic $n$ is defined as a sinusoidal wave of amplitude $V_n$ and relative phase $\phi_n$ with respect to that of the third harmonic, resulting in an applied driving voltage of
\begin{equation}\label{eq:drivsig}
\begin{split}
    V\left(t\right) = V_\mathrm{dc} + V_1 \sin(2\pi f_gt + \phi_1) + V_2 \sin(4\pi f_g t + \phi_2) + \\
    V_3 \sin(6\pi  f_g t).
\end{split}
\end{equation}
The optimization problem consists in determining the relative amplitudes and phases of the different harmonics in order to minimize $\delta$.  The simplified generic RRM model described in \cite{Sacher_ox_08_v16_p15741} was used to numerically optimize the synthesized comb in conjunction with the DE algorithm. This optimization was performed offline.

In the optimization of the RRM EOFC, the methodology applied in \cite{Pinto_ox_21_v29_p23447} to the optimization of gain-switched laser combs was followed.  Its goal was to determine the best values of the amplitudes $V_n$ and relative phases $\phi_n$ of the input signal described by \eqref{eq:drivsig}, as well as the fundamental frequency $f_g$ and the detuning $\Delta \lambda$. Since the goal is to minimize $\delta$, the fitness function  $f\left(\mathbf{P}\right)$ used in the optimization was defined as the unbiased sample variance of the line power levels in the normalized comb spectrum when considering a fixed target number of lines $N_l$, i.e.
\begin{equation} \label{eq:var}
 f\left(\mathbf{P}\right) = \frac{1}{N_l - 1} \sum_{i=1}^{N_l}\left( P_i - \mu\right)^2,
\end{equation}
where $\textbf{P} = \left(P_i\right)$, $i = 1\ldots N_l$ is the vector of relative power levels of the $N_l$ lines in the comb, and $\mu$ its sample mean. The optimization method is further described in the pseudo-code of Algorithm~1.

\begin{algorithm}[t]
\SetAlgoLined
 Initialize $N_{c}$ solutions $\mathbf{x}_{1:N_c}$\;\\
 Find best initial solution $\mathbf{g}_{best}$ by f($\mathbf{x}_{1:N_c}$)\;\\
 \While{$k~\mathrm{less~than}~N_i$}{
 \For{$i=1~\mathrm{to}~N_c$}{
  Generate donor vector $\mathbf{v}_{k,i}$\;\\
  Generate trial vector $\mathbf{u}_{k,i}$\;\\
  Check RRM EOFC solutions limits\;\\
  Apply $\mathbf{u}_{k,i}$ to RRM EOFC\;\\
  Evaluate RRM EOFC flatness by $f(\mathbf{u}_{k,i})$\;\\
   \If{f($\mathbf{u}_{k,i}$) $<$ f($\mathbf{x}_{k-1,i}$)}{
   $\mathbf{x}_{k,i}$ = $\mathbf{u}_{k,i}$\;
   }
   \Else{
   $\mathbf{x}_{k,i}$ = $\mathbf{x}_{k-1,i}$\;}
   {
   \If{f($\mathbf{u}_{k,i}$) $<$ f($\mathbf{g}_{best})$}{
   $\mathbf{g}_{best}$ = $\mathbf{u}_{k,i}$\;
   }
   }
  }
 }
\caption{Differential evolution algorithm for the flatness optimisation of RRM-generated EOFCs. $N_i$ is the number of iterations and $N_c$ is the number of solutions.}  \label{alg:de}
\end{algorithm}

\begin{figure}[t]\label{fig:numresults7lines}
    \centering
    \subfigure[]{\includegraphics[scale=0.28]{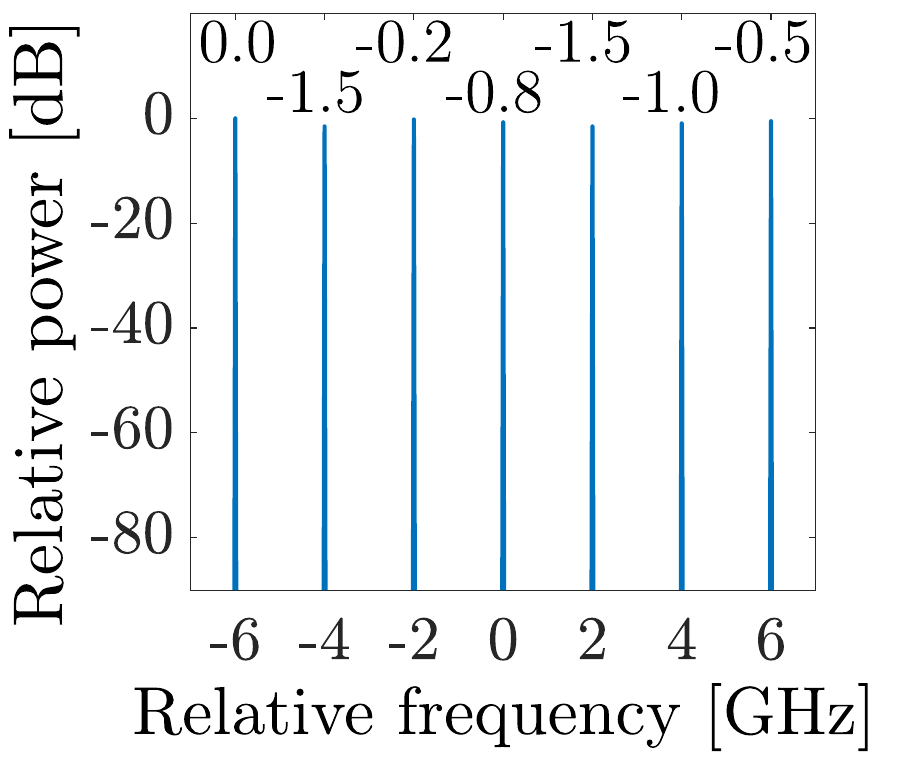}
        \label{fig:numresults7lines_spectrum}}
    \subfigure[]{\includegraphics[scale=0.28]{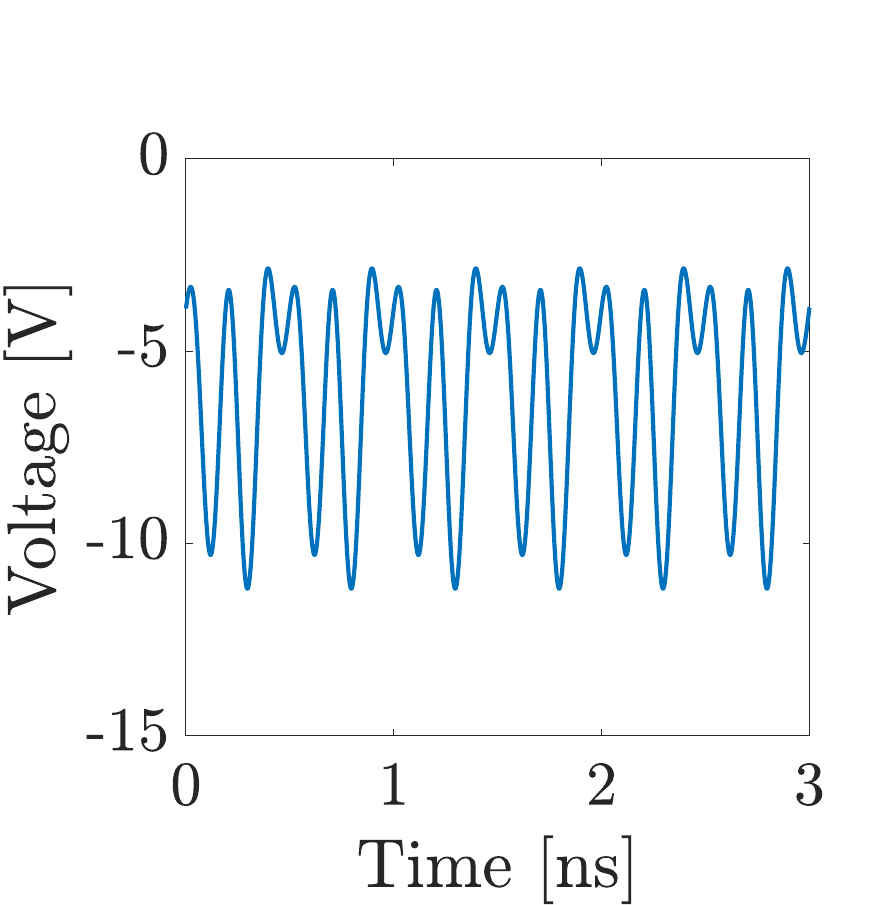}
        \label{fig:numresults7lines_driving_signal}}
     \caption{RRM comb optimization for 7 lines. (a)~Optimized EOFC. (b)~Optimized RRM driving signal.}
\end{figure}

To perform the optimization, the total driving voltage was constrained to the range $-14$~V to 0~V due to the physical limits of the phase-shifter of the RRM. The fundamental frequency $f_g$ was restricted to the interval between 1~GHz and 10~GHz and ~the detuning $\Delta \lambda$ was allowed to vary between $-200$~pm and 200~pm. The target number of comb lines $N_l$ was fixed while all aforementioned parameters were allowed to vary in the specified ranges. Optimizing the overall spectra over different values of $N_l=3,5,7,9$, the best results were obtained for a 7-line comb with a power imbalance of $\delta = 1.5$~dB, as shown in Fig.~\ref{fig:numresults7lines_spectrum}. The driving signal used to achieve this optimal comb flatness is represented in Fig.~\ref{fig:numresults7lines_driving_signal}. The corresponding values of the optimization parameters are listed in Table~\ref{tab:resultsdeoptm}.

\begin{table}
    \centering
    \begin{tabular}{|c||>{\centering\arraybackslash}p{1cm}|>{\centering\arraybackslash}p{1cm}||>{\centering\arraybackslash}p{1cm}|>{\centering\arraybackslash}p{1cm}|}
    \hline
    Parameter & Sim. (7~lines) & Sim. (9~lines) & Exp. (7~lines) & Exp. (9~lines)\\ \hline
    $V_\mathrm{dc}$ [Volts] &-6.2 & -5.3 & -7.0 & -7.0 \\ \hline
    $V_{1}$ [Volts] & 1.8 & 2.6 & 1.4 & 1.5 \\ \hline
    $V_{2}$ [Volts] & 2 & 2.7 & 1.5 & 1.5 \\ \hline
    $V_{3}$ [Volts] & 2.6 & 2.4 & 2.0 & 1.4 \\ \hline
    $\phi_{1}$ [rad] & 2.0 & 3.0 & 2.0 & 3.0 \\ \hline
    $\phi_{2}$ [rad] & 2.8 & 1.6 & 2.8 & 1.6 \\ \hline
    $f_g$ [GHz] & 2.0 & 1.0 & 0.5 & 0.5 \\ \hline
    $\Delta \lambda$ [pm] & -5 & -28 & +22 & +30 \\ \hline
    $\delta$ [dB] & 1.5 & 6.8 & 2.9 & 5.4 \\
   \hline
    \end{tabular}
    \caption{Optimum values of the optimization parameters obtained by the DE algorithm used in conjunction with the RRM numerical model ("Sim.") and after experimental optimization ("Exp."). In both cases, optimizations were performed for target numbers of lines of 7 and 9.}
    \label{tab:resultsdeoptm}
\end{table}

\begin{figure}[t]
    \centering
    \subfigure[]{\includegraphics[scale=0.27]{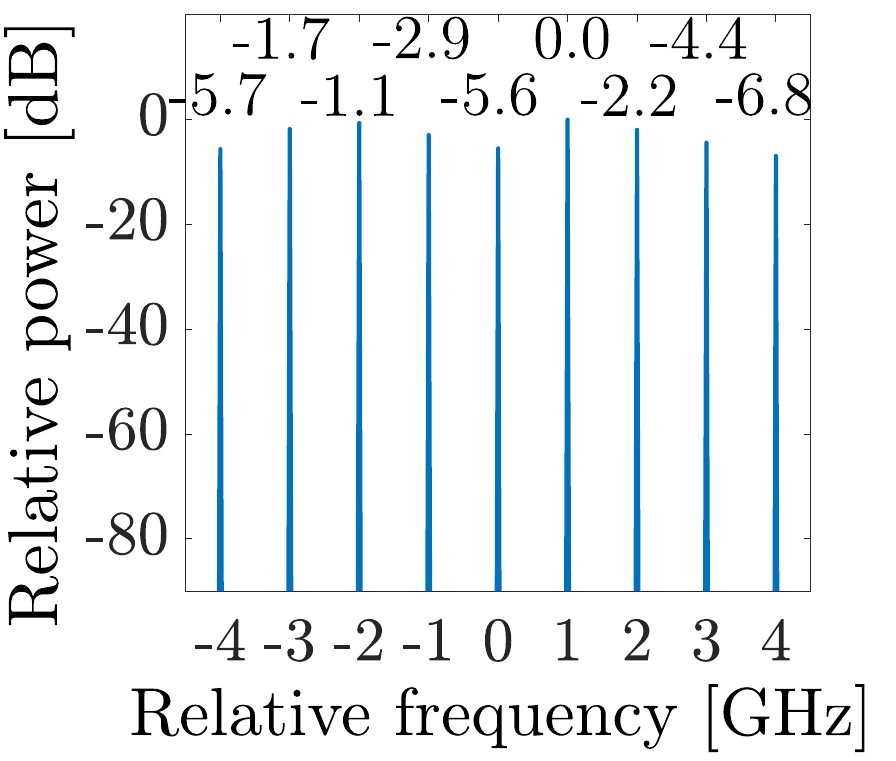}
        \label{fig:numresults9lines_spectrum}}
    \subfigure[]{
        \includegraphics[scale=0.27]{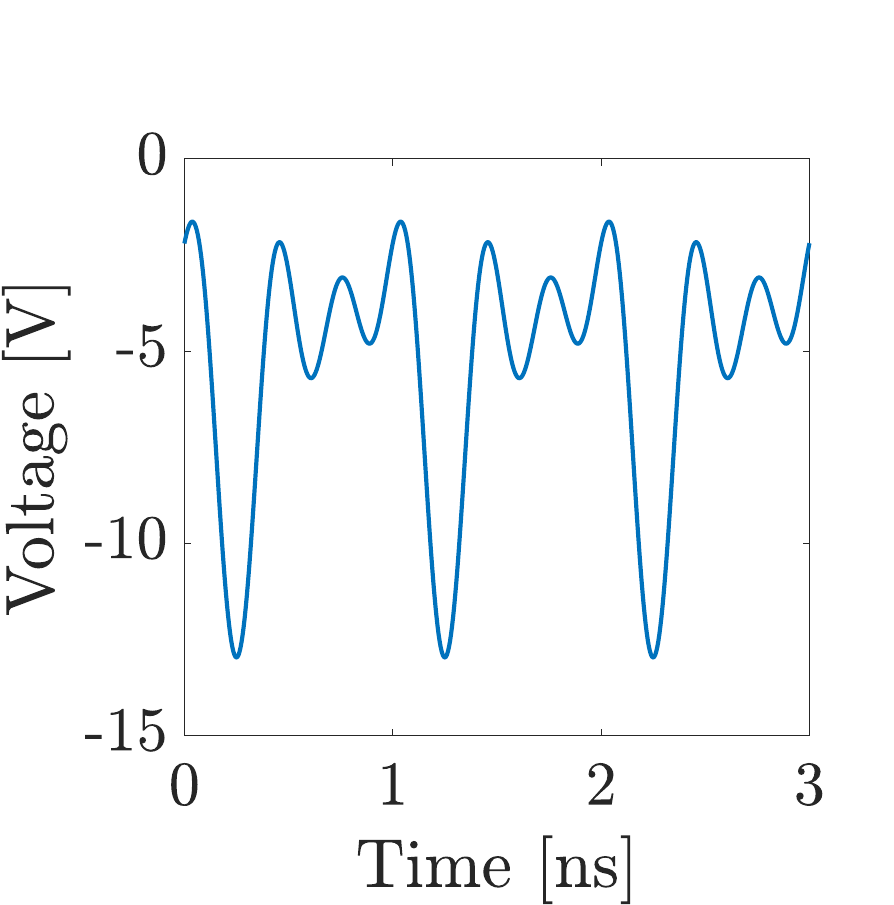}
        \label{fig:numresults9lines_driving_signal}
    }
    \caption{RRM comb optimization for 9 lines. (a)~Optimized EOFC. (b)~Optimized RRM driving signal.}
    \label{fig:numresults9lines}
\end{figure}

For a target 9-line comb, a power imbalance of $\delta= 6.8$ dB was achieved, as illustrated in the spectrum of Fig.~\ref{fig:numresults9lines_spectrum}, from the driving signal of Fig.~\ref{fig:numresults9lines_spectrum} obtained with the optimized parameters listed in Table~\ref{tab:resultsdeoptm}. Even though $\delta$ is larger than in the optimized 7-line comb, the flatness is still improved compared to the single harmonic case of Fig.~\ref{fig:mes_sinusoidal_5lines}, confirming the efficiency of the optimization method.

\begin{figure}
    \centering
    \subfigure[]
    {
        \includegraphics[scale=0.28]{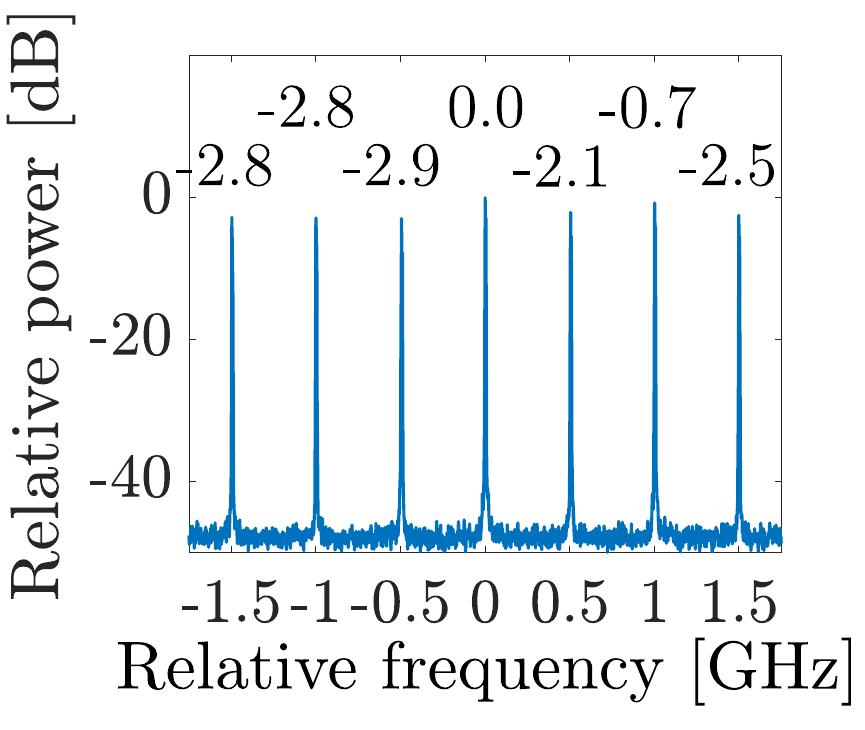}
        \label{fig:mes_7lines_spectrum}
    }
    \subfigure[]
    {
        \includegraphics[scale=0.28]{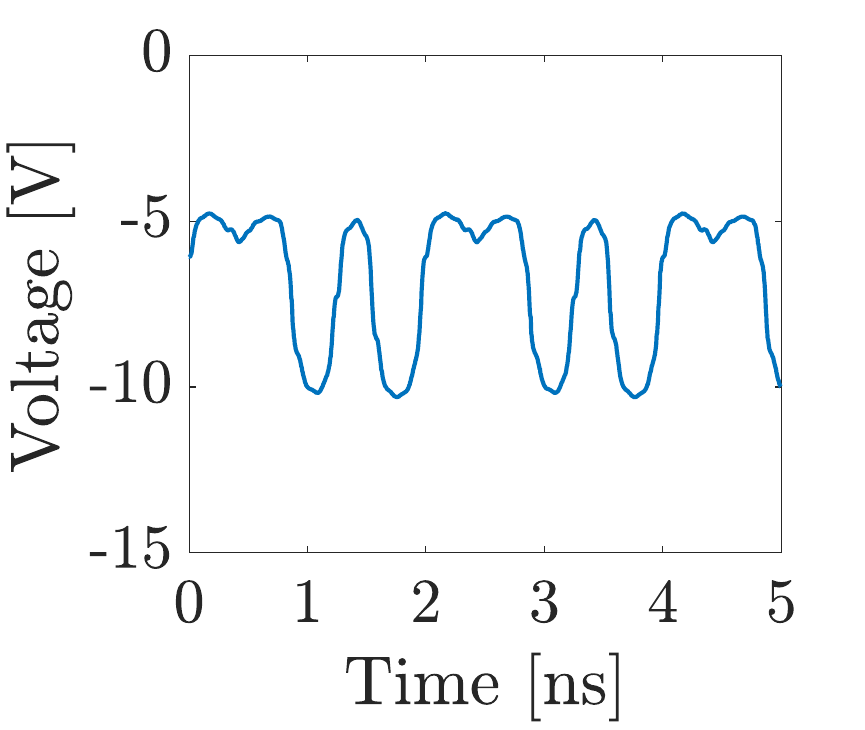}
        \label{fig:mes_7lines_signal}
    }
    \subfigure[]
    {
        \includegraphics[scale=0.28]{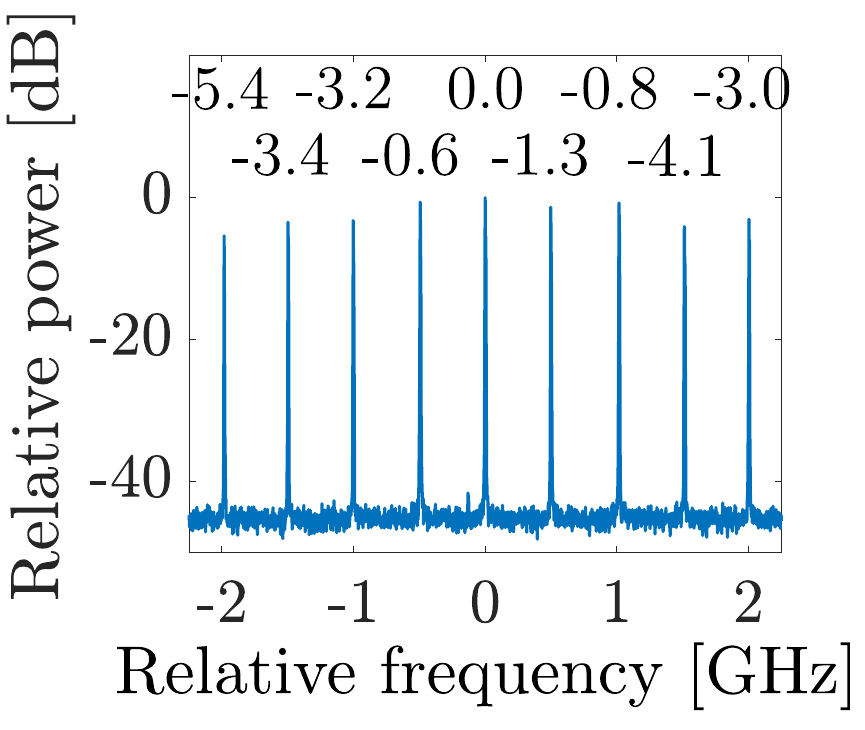}
        \label{fig:mes_9lines_spectrum}
    }
    \subfigure[]
    {
        \includegraphics[scale=0.28]{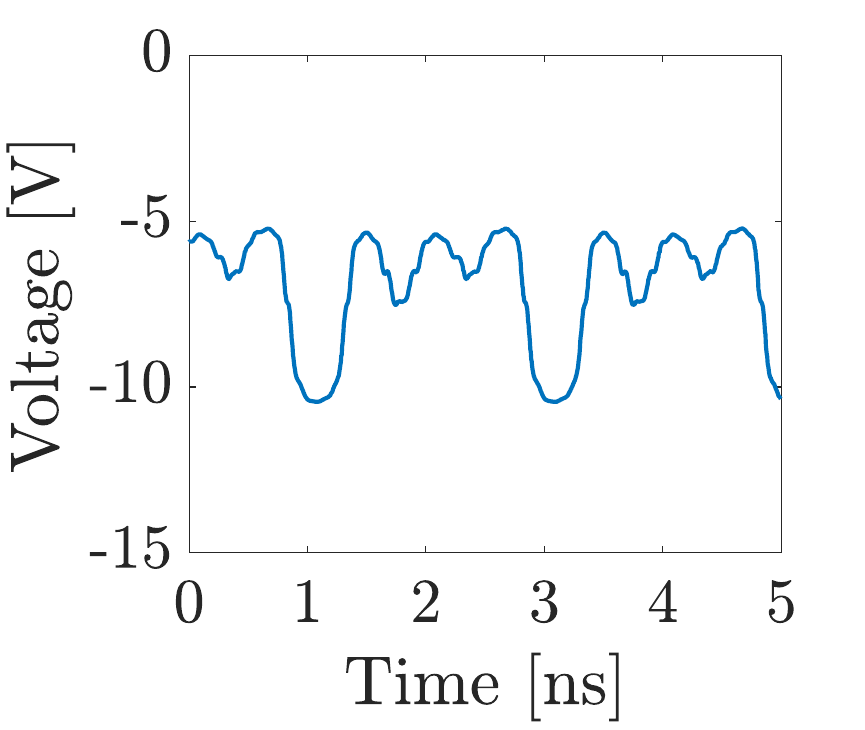}
        \label{fig:mes_9lines_signal}
    }
    \caption{Measured optimized spectra and their corresponding driving voltage signals for EOFCs containing 7 and 9 lines. (a) and (b) represent the output spectrum and the applied voltage as a function of time for the 7-line EOFC ,respectively. (c) and (d) represent the output spectrum and the applied voltage as a function of time for the 9-line EOFC, respectively.}
\end{figure}

In the experimental validation of the optimization procedure, the electrical driving signals resulting from the numerical optimization were synthesized using an arbitrary waveform generator with $10-$bit resolution at a sampling frequency of 24~GS/s. The generated waveforms have a maximum peak-to-peak amplitude of 1~V. An RF driver amplifier was then used to reach a peak-to-peak amplitude of up to 6.4~V. Due to driver amplifier limitations, this peak-to-peak voltage is smaller than what was established in the numerical optimization phase, where the optimum driving signals had peak-to-peak voltage values of 8.3~V and 11.4~V for target combs with 7 and 9 lines, respectively. The corresponding amplitudes for each harmonic are given in Table~\ref{tab:resultsdeoptm} as experimental parameters. A further discrepancy between the numerical and experimental optimizations stem from limitations of the RRM model. Indeed, the model assumes a phase-shifter made from an ideal abrupt PN junction distributed over the entire circumference of the ring. Furthermore, no electrical response, which can be compensated by adjusting the relative amplitude of the harmonics, is included. Nevertheless, the point here is not to obtain a perfect quantitative match between the numerical and experimental results, which would require a more accurate RRM model, but to demonstrate qualitatively that the proposed optimization procedure can be applied to experimental comb synthesis. The calculated comb spectra may therefore differ from the measured one. Due to the aforementioned limits of the numerical model, the wavelength detuning was fine-tuned manually while the applied driving voltage was kept unchanged, with the parameters listed in Table~\ref{tab:resultsdeoptm}. In order to circumvent the effect of the actual electrical response of the fabricated RRM, which was ignored in the simulations, the fundamental frequency was kept at a lower value of $f_g = 500$~MHz, where this response has little impact given that only 3 harmonics are considered. This is not a fundamental limitation to the application of our optimization procedure to higher fundamental frequencies, but the electrical response would need to be properly compensated by adjusting the amplitudes and relative phases of the harmonics. An optimal 7-line comb with a flatness of $\delta=2.9$~dB was measured, as illustrated in Fig.~\ref{fig:mes_7lines_spectrum}, together with the applied driving signal in Fig.~\ref{fig:mes_7lines_signal}. In spite of the model limitations, the experimentally measured imbalance is only 1.4~dB higher than the one obtained with the simplified model through numerical optimization. The same process was repeated towards the generation of a 9-line comb, with a measured flatness of $\delta=5.4$~dB, as shown in Fig.~\ref{fig:mes_9lines_spectrum}. The discrepancy with the numerical results is, once again, the result of the relative simplicity of the numerical model, which was not designed for accurate performance prediction of a real fabricated device. In particular, the well-known trade-off between bandwidth and modulation amplitude \cite{Yu_ox_14_v22_p15178} (experimentally, electro-optical 3-dB bandwidths ranging from 4~GHz with large modulation amplitude to 25~GHz with small modulation amplitude were measured for detunings varying between 0 and 60~pm for the RRM used in this work) is not captured by the model. In any case, experimental 7- and 9-line combs with optimum flatness were generated from a single RRM with a 3-harmonic driving signal, thus demonstrating the efficiency of the optimization procedure.

In summary, optimized EOFCs have been generated using a single silicon RRM with harmonic superposition in the driving signal. Thanks to the differential evolutionary algorithm coupled with a numerical model describing the RRM, optimal driving signals have been synthesized to generate flat EOFCs containing 7 and 9 lines. Experimentally, flatness values of 2.9~dB in 7 lines and 5.4~dB in 9 lines have been obtained, which correspond to the order of magnitude expected from numerical optimization with a simplified RRM model. These results should ease flat EOFC generation for spectroscopy applications at the current frequency spacing (500~MHz), but also for optical communication applications using the same method with RRMs designed to maximize optical and electrical bandwidths, which should allow to reach the same flatness with modulation frequencies of tens of GHz.

\begin{backmatter}
\bmsection{Funding} Région Bretagne (ARED); Lannion-Trégor Communauté (SiTrans project); H2020 Marie Skłodowska-Curie Actions (814276; 754462); European Research Council (ERC-CoG FRECOM project grant agreement no. 771878); Villum Fonden (VYI OPTIC-AI grant no. 29344)

\bmsection{Disclosures} The authors declare no conflicts of interest.

\bmsection{Data availability} Data underlying the results presented in this paper are not publicly available at this time but may be obtained from the authors upon reasonable request.

\end{backmatter}

\bibliography{references}

\bibliographyfullrefs{references}

\end{document}